# Supergravity Interactions In The SU(4)xU(1) Vacuum Of Global Supersymmetry

J. Towe

Department of Physics, The Antelope Valley College, Lancaster, CA 93536-5426
TEL 661 722 6427, FAX 661 722 6416 jtowe@avc.edu

## Abstract

It is shown that quarks and leptons can couple intrinsically to pure supergravity, preserving for this context the degeneracy of globally supersymmetric vacua, and specifically the vacuum that is characterized by SU(4)XU(1). A fundamental representation of SUSY SU(4)XU(1) is constructed in terms of quarks, and it is demonstrated that the symmetry of this fundamental representation is invariant under supergravity interactions that produce quark-lepton transitions. The importance of this tensorial flavor structure is that it predicts a new, seventh quark of charge -1/3, a left-handed (non-strange) version of the strange quark. Among the observable consequences of this prediction are non-strange versions of the strange hadrons.

# INTRODUCTION

Global supersymmetry is associated with three vacua: one characterized by SU(5); a second by SU(4)XU(1) and a third by SU(3)XSU(2)XU(1). Current theory removes this degeneracy of vacua by permitting a minimal breaking of supersymmetry in the hidden sector. It is preferred that the vacuum corresponding to the standard model be directly selected. But one can demonstrate a context in which quarks and leptons intrinsically couple to pure supergravity. In this context, the absence of a hidden sector preserves the degeneracy of vacua that includes the vacuum of symmetry SU(4)XU(1). Moreover, one can construct a fundamental representation of SUSY SU(4)XU(1) in terms of quarks, and demonstrate that the symmetry of this fundamental representation is invariant under pure supergravity interactions that produce quark-lepton transitions.

The proposed fundamental representation of SUSY SU(4)XU(1) can be constructed only after some qualifying remarks. The quarks and leptons under discussion are massless. It is therefore assumed that there is no rest frame, so that helicity is regarded as an intrinsic particle property. In this context, the SUSY fundamental representation of SU(4)XU(1) is given as $\{4,1,\bar{1},1\}$ (where 1 represents a quark, $\bar{1}$ represents an antiquark, $\{1, \bar{1}, 1\}$ represents a triplet of net spin 3/2 and 4 represents a spin-2 composite of four quarks of like helicity). This fundamental representation either requires color or more flavors than those contained in the three quark generations. Color is precluded because the symmetry SUSY SU(4)XU(1) is preserved under supergravity interactions that produce quark-lepton transitions, and because leptons involve no color. In this context the required quark flavor is a left-handed generational partner of the charmed quark. Usually one would hesitate to introduce a new quark flavor because this would seem to violate six-quark—six-lepton symmetry. But one can accomplish what is required here by choosing the new quark as a left-handed, non-strange version of the strange quark—related to the strange as the left-handed electron is related to the right-handed electron. By this choice one avoids violation of six-quark—six-lepton symmetry and obtains three fundamental representations of SUSY SU(4)XU(1).

The existence of the predicted new quark would cure an anomaly that exists in the traditional theory of generations in which up and top quarks are complemented by left-handed generational partners (the down and bottom quarks) while the charmed quark is complemented by a right-handed generational partner--the strange quark. Moreover, the predicted quark indicates non-strange versions of the strange hadrons.

# I. A Fundamental Representation Of Susy SU(4)XU(1)

A fundamental representation of SU(4)XU(1) is provided by

$$\{\mathbf{4}, \mathbf{1}\}, \qquad (1)$$

but in the high energy, relativistic context described above, SUSY SU(4)XU(1) requires a fundamental representation

$$\{\mathbf{4}, \mathbf{1}, \bar{\mathbf{1}}, \mathbf{1}\}, \qquad (2)$$

where $\bar{\mathbf{1}}$ refers to an anti-quark, where $\{\mathbf{1}, \bar{\mathbf{1}}, \mathbf{1}\}$ refers to a triplet of quarks in an asymptotic limit where they behave as individual particles and where **4** refers to a composite of 4 quarks. To qualify as a fundamental representation of SUSY SU(4)XU(1), one of three options must hold: either the triplet is of spin ½ and the **4** of spin 1; the triplet is of spin ½ and the **4** is of spin zero; or the triplet is of spin 3/2 and the **4** is of spin two. This discussion will focus upon the third option. In this case, $\mathbf{1}, \bar{\mathbf{1}}$ and **1** must be of the same helicity (the triplet $\{\mathbf{1}, \bar{\mathbf{1}}, \mathbf{1}\}$ is of spin 3/2), and the composite **4** must consist of four fermions of the same helicity (the **4** is of spin 2).

    The fundamental representation (2) either requires color or more flavors than those provided by SU(5) [D. Nordstrom, 1992]. (One may question how the generations are distinguished before the fermions are endowed with mass. This distinction is based upon a postulated generational quantum number: In Figures 1A and 1B, the off-plane axis is characterized as a generational axis; and each of the depicted particles is characterized by a quantum number associated with the generational axis as well as quantum numbers associated with the $I_3$ and Y axes.) Color cannot be enlisted because interactions preserving SUSYXSU(4)XU(1) produce quark-lepton transitions, and leptons are not associated with color. Additional generations of flavor are therefore required. Even in the context of the three known flavor generations however, the representation $\{\mathbf{4}, \mathbf{1}, \bar{\mathbf{1}}, \mathbf{1}\}$ cannot be realized unless a new quark is introduced. The incomplete picture is as follows:

$$\begin{pmatrix} c \\ ? \\ T \\ B \end{pmatrix}, u, d, \bar{s}; \quad \begin{pmatrix} u \\ d \\ T \\ B \end{pmatrix}, c, ?, \bar{s}; \quad \begin{pmatrix} u \\ d \\ c \\ ? \end{pmatrix}, T, B, \bar{s} \qquad (3)$$

(where $\bar{s}$ represents an anti-strange quark, a realization of $\bar{\mathbf{1}}$). One is reluctant, of course, to suggest this option since it appears to challenge the currently accepted six-quark—six-lepton symmetry. However, from (3) above we see that the required quark is a left-handed generational partner of the charmed quark, so that what is needed can be accomplished by choosing the new quark as a left-handed version of the strange quark—related to the strange as the left-handed electron is related to the right-handed electron. In this context, the violation of six-quark--six-lepton symmetry is avoided and three fermionic realizations of {**4**, **1**, **$\bar{1}$**, **1**} are obtained:

$$\begin{pmatrix} c \\ 7 \\ T \\ B \end{pmatrix}, \text{u, d and } \bar{s}. \tag{4}$$

$$\begin{pmatrix} u \\ d \\ T \\ B \end{pmatrix}, \text{c, 7 and } \bar{s} \tag{5}$$

and

$$\begin{pmatrix} u \\ d \\ c \\ 7 \end{pmatrix}, \text{T, B and } \bar{s}. \tag{6}$$

Note that the composite **4**'s of (4), (5) and (6) are of spin-2 (consisting exclusively of left-handed fermions); and that the quark triplets cumulatively represent multiplets of spin-(3/2), so that (4), (5) and (6) are supersymmetric multiplets [J. Towe, 1997]. Each 4-tuplet is composed of four left-handed quarks, and therefore carries a color C (chosen from red, yellow or blue), a charge of 2/3 and a left-handed helicity. Note that the strange and anti-strange quarks are common to all three triplet generations and that the supersymmetric flavor generations described by (4), (5) and (6) can be represented as three orientations about the hypercharge axis of the Figure 1A configuration. The 4-tuplet

$$\begin{pmatrix} c \\ 7 \\ T \\ B \end{pmatrix} \tag{7}$$

corresponds to the four vertices of Figure 1A that lie off the ($I_3$, Y)-plane and above the (G, $I_3$)-plane; while the anti-particles of c, 7, T and B correspond to positions depicted off the ($I_3$, Y)-plane and below the (G, $I_3$)-plane). However, if the Figure 1A configuration is rotated through an angle of $2\pi/3$ about the Y-axis (hypercharge axis) of Figure 1A, then the off-plane particles c, 7, T, B which constituted the relevant 4-tuplet before the rotation, are replaced by u, d, c, 7; and the triplet on the $(I_3,Y)$-plane becomes T, B and $\bar{s}$. An additional rotation (through $2\pi/3$) produces a 4-composite of off-plane quarks (u, d, T, B) and three on-plane quarks: c, 7 and $\bar{s}$. The optional orientations about Y are clearly permitted by the degree of freedom which lies off the $(I_3, Y)$-plane of Figure 1A. The axis corresponding to this degree of freedom is therefore designated "G" with reference to "generation."

## II. LOCALLY SUPERSYMMETRIC INTERACTIONS

The action of global supersymmetry can be gauged in the Noether tradition to produce an action for local supersymmetry. This local action (the action of pure supergravity) on spacetime is given by

$$S = -(1/2\kappa^2) \int d^4x\, |\det e|\, R - (1/2) \int d^4x\, \varepsilon^{\mu\nu\rho\sigma}\, \bar{\psi}_\mu \gamma_5 \gamma_\nu \bar{D}_\rho \Psi_\sigma, \qquad (8)$$

which is the sum of the Einstein action (R is the curvature scaler) and the action of the Rarita-Schwinger field, which is covariantized according to the covariant derivative

$$D_\mu = \partial_\mu - i\, \bar{\omega}_{\mu mn}(\sigma^{mn}/4), \qquad (9)$$

where

$$\bar{\omega}_{\mu mn} = \omega_{\mu mn} + (i\kappa^2/4)(\bar{\psi}_\mu \gamma_m \psi_n + \bar{\psi}_m \gamma_\mu \psi_n - \bar{\psi}_\mu \gamma_n \psi_m)$$

and

$$\omega_{\mu mn} = (1/2)e_m{}^\nu(\partial_\mu e_{n\nu} - \partial_\nu e_{n\mu}) + (1/2)(e_m{}^\rho e_n{}^\sigma \partial_\sigma e_{\rho p}\, e_\mu{}^p) - (m \leftrightarrow n),$$

where $\omega_{\mu mn}$ is the spin connection. (The minimal covariant derivative of general relativity is augmented here by the term quadratic in the Rarita-Schwinger field to achieve invariance of the action at order $\kappa^2$.)

The action (8) requires a graviton vertex operator, consisting of a right-moving boson and a left-moving fermion, which interface with a spin-(3/2) field (the gauge field of local supersymmetry), thereby inducing the spin-(3/2) field to produce a spin-2 field [D. Bailin, 1994]. (Note that above described vertex operator is precisely the same operator that arises in the theory of closed strings from the right-moving Neveu-Schwarz ground state and the left-moving Ramond ground state.)

Note (Figures 2, 3 and 4) that the supersymmetric status of the vertex can be maintained if the in-moving gravitino is replaced with an out-moving fermion-boson pair of like helicity. This is a helicity opposite that which would characterize the in-moving gravitino. The remarkable thing about this picture is that if the in-moving fermion is a quark and the out-moving fermion is a lepton, or vice versa, then the out-moving spin-2 field is always either right-handed of charge -2/3, or left-handed, of charge +2/3. This supersymmetric substitute for the usual graviton vertex operator will serve as the tree-level proto-type for the locally supersymmetric interactions to which we have referred. An example of this kind of interaction is depicted in Figure 2, where the above described vertex operator involves a right-handed anti-quark of charge 1/3 and anti-color $\overline{C}$, and a right-handed boson of charge zero. The result of this interaction is (as described above) the release of a right-handed spin-2 (anti) field of charge -2/3 and anti-color $\overline{C}$. Specific interactions involving the same graviton vertex operator are described below.

In the interaction depicted by Figure 2, a left-handed lepton of charge –1 absorbs a right-handed boson of charge zero. This combination simultaneously radiates a spin-2 field of right-handed helicity, charge –2/3 and a single anti-color (anti-red, anti-yellow or anti-blue) which is identified simply as "$\overline{C}$", a left-handed boson of charge zero and a left-handed fermion of charge –1/3, which is characterized by the color C (anti-color of $\overline{C}$). Due to the nature of its quantum numbers, this particle is associated with the down quark. Secondly, the spin-2 **4** absorbs a left-handed boson of charge zero, and a particle identical to that which is (due to its quantum numbers) identified as the down quark, and immediately radiates a right-handed boson of charge zero and a left-handed fermion of charge –1 which, due to its quantum numbers, is identify as an LH electron.

In the interaction depicted by Figure 3, a left-handed fermion of charge zero (which I identify as the LH electron's neutrino) absorbs a right-handed boson of charge zero. This combination simultaneously radiates a spin-2 anti-field of right-handed helicity, charge –2/3 and anti-color $\overline{C}$, a left-handed boson of charge zero and a left-handed fermion, characterized by a charge of 2/3 and the color C, which I identify as an up quark. Secondly, the spin-2 anti-field simultaneously absorbs a left-handed boson of charge zero and a left-handed fermion of charge 2/3 and color C, which I have identified as an up quark, and radiates a right-handed boson of charge zero and left-handed fermion of charge zero, which I have identified as the LH electron's neutrino.

The interaction depicted by Figure 4 is an opposite-handed version of the interactions that are depicted by Figures 2 and 3. In Figure 4, a right-handed quark of charge –1/3 and color C absorbs a left-handed boson of charge zero and simultaneously radiates a left-handed spin-2 field of charge 2/3 and color C, a right-handed boson of charge zero and a right-handed fermion of charge –1, which, due to its quantum numbers, is identified as an RH electron. Secondly, the spin-2 field absorbs a right-handed boson of charge zero and a right-handed fermion of charge –1 (which I identify as an RH electron) and radiates a left-handed boson of charge zero and a right handed fermion of charge –1/3 and color C, which I identify as a strange quark.

Thus, the interactions depicted by Figures 2, 3 and 4 produce leptonic realizations

$$\begin{pmatrix} \mu^-_L \\ \nu^{\mu-} \\ \tau^-_L \\ \nu^{\tau-} \end{pmatrix}, e^+_L, e^-_L \text{ and } \nu^{e-} \qquad (10)$$

($e^+_L$ is the anti-particle of the right-handed electron),

$$\begin{pmatrix} e^-{}_L \\ \nu^{e-} \\ \mu^-{}_L \\ \nu^{\mu-} \end{pmatrix}, e^+{}_L, \tau^-{}_L, \nu^{\tau-} \qquad (11)$$

and

$$\begin{pmatrix} e^-{}_L \\ \nu^{e-} \\ \tau^-{}_L \\ \nu^{\tau-} \end{pmatrix}, e^+{}_L, \mu^-{}_L, \nu^{\mu-} \qquad (12)$$

of {**4**, **1**, $\bar{\mathbf{1}}$, **1**}, which are depicted by Figure 1B. Clearly, the locally supersymmetric interactions that are depicted by Figures 2, 3 and 4 preserve the flavor supersymmetry that is characteristic of {**4**, **1**, $\bar{\mathbf{1}}$, **1**} and is illustrated by Figures 1A and 1B. Thus, the proposed fermionic fundamental representation of SUSY SU(4)XU(1); i.e. the proposed fermionic realization of {**4**, **1**, $\bar{\mathbf{1}}$, **1**} attains a tensorial status in the theory of pure supergravity. Incidentally the spin-2 fields which correspond to the fermionic **4**'s mediate transitions from quark-leptonic pairs to the isospin conjugates of those pairs.

## CONCLUSION

Global supersymmetry is associated with three vacua: one characterized by SU(5); a second by SU(4)XU(1) and a third by SU(3)XSU(2)XU(1). Most current theories avoid the vacuum that associates with SU(4)XU(1). The degeneracy of vacua is removed by the breaking of supersymmetry in the hidden sector. It is preferred that the vacuum corresponding to the standard model be directly selected. But this discussion demonstrated a context in which quarks and leptons intrinsically couple to pure supergravity. In this context, the absence of a hidden sector preserves the degeneracy of vacua that includes the vacuum of symmetry SU(4)XU(1). Moreover, we constructed a fundamental representation {4,1,$\bar{1}$, 1} of SUSY SU(4)XU(1) in terms of fermions and demonstrated that the symmetry of this fundamental representation is invariant under the pure supergravity interactions that produce quark-lepton transitions.

The importance of the proposed tensorial flavor structure is that it predicts a new, seventh quark--a left-handed, non-strange version of the strange quark). The prediction of this new quark cures an anomaly that exists in the traditional theory of generations in which up and top quarks are complemented by left-handed generational partners (the down and bottom quarks) while the charmed quark is complemented by a right-handed generational partner--the strange quark. Moreover, the predicted quark indicates non-strange versions of the strange hadrons (e.g. of the recently observed D-meson [BABAR Collaboration, 2003]). Additional consequences of the proposed theory are:

1. An explanation of the fermionic generations (these were a necessary condition for the construction of the proposed fermionic fundamental representations of SU(4)XU(1));

and

2. Locally supersymmetric interactions which do not include the usual superpartners.

# FIGURES

1A. Flavor Supersymmetry of Quarks

1B. Flavor Supersymmetry of Leptons

2. The Transition from a Down Quark to an Electron

3. The Transition from an Up Quark to the Electron's Neutrino

4. The Transition from a Strange Quark to a Right-Handed Electron

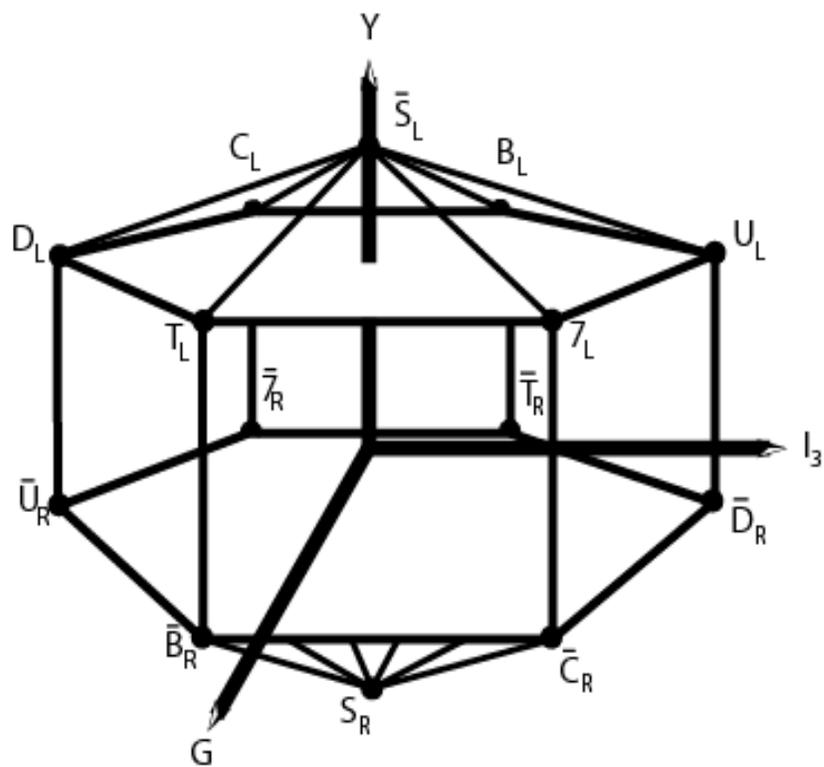

Figure 1A

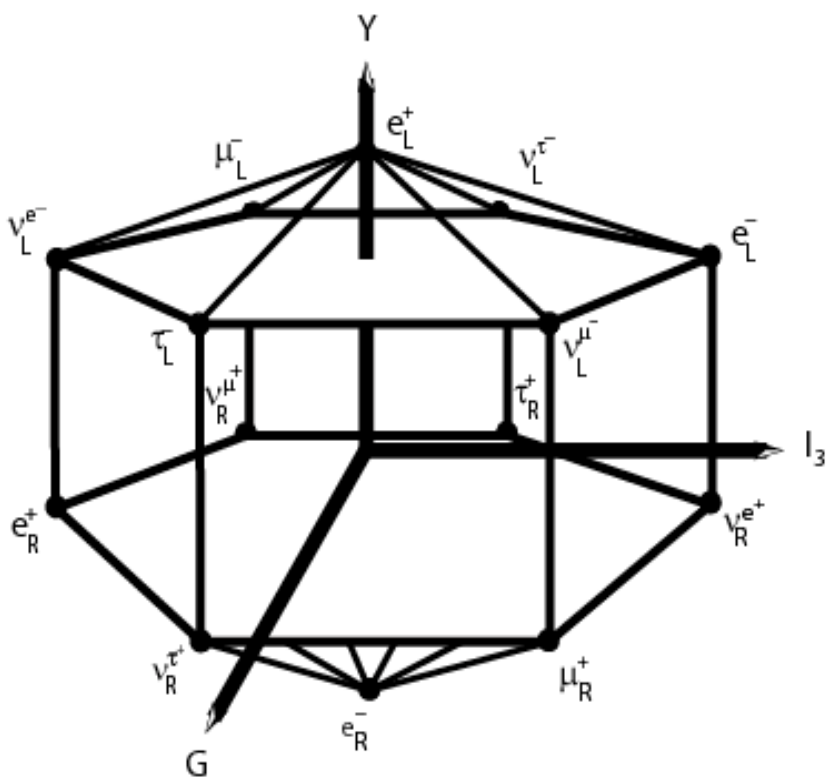

Figure 1B

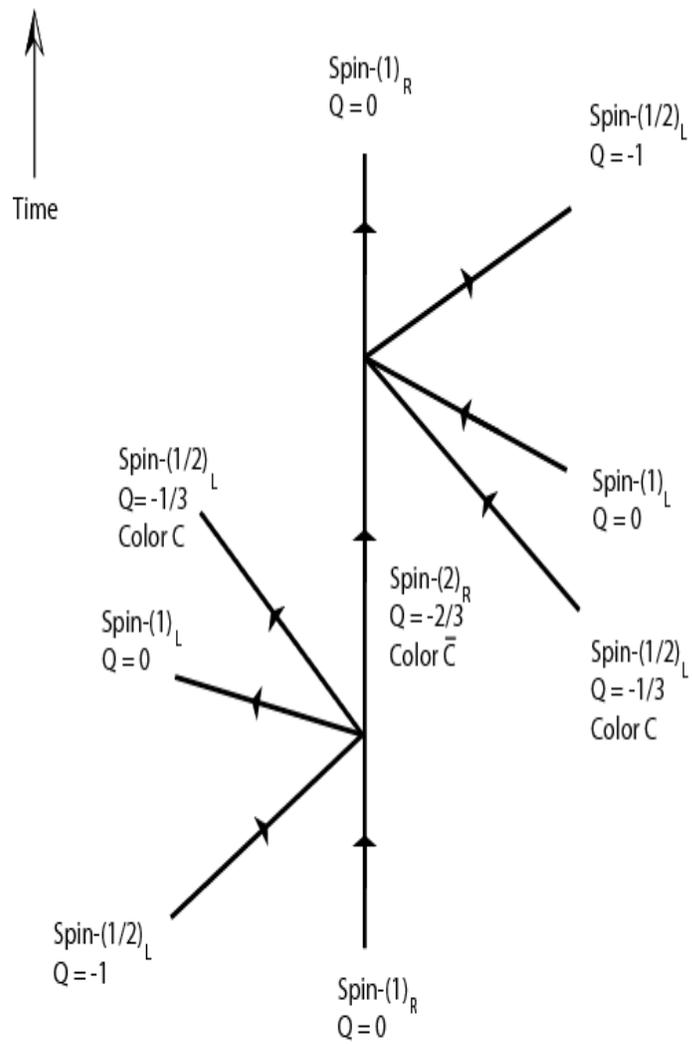

Figure 2

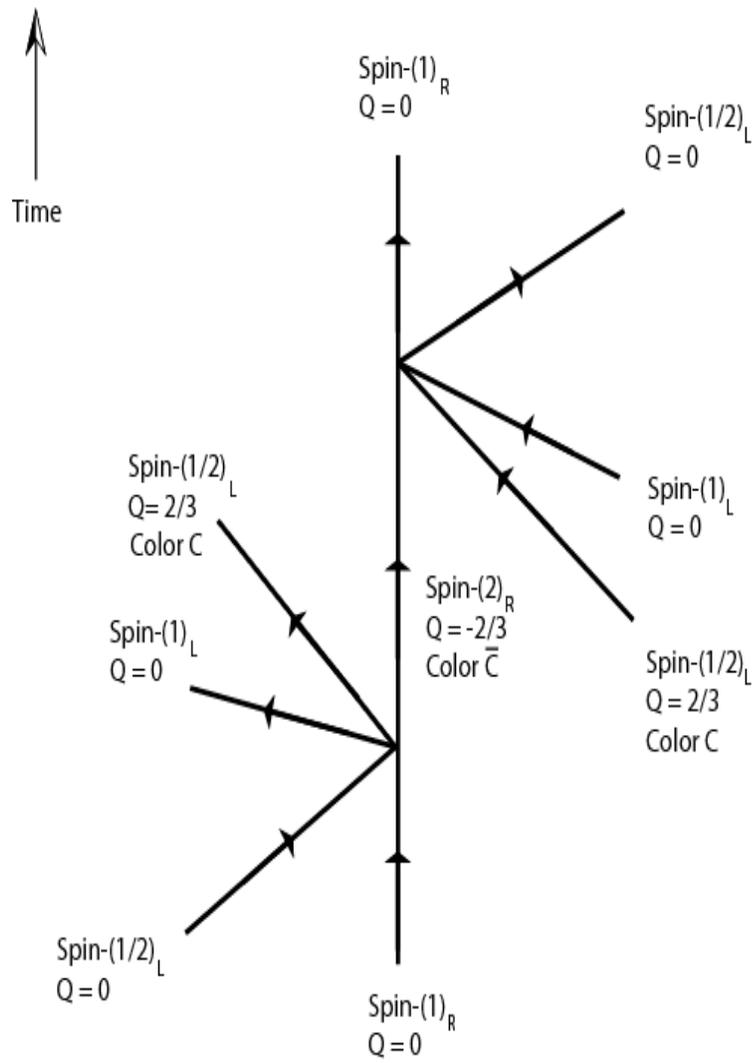

Figure 3

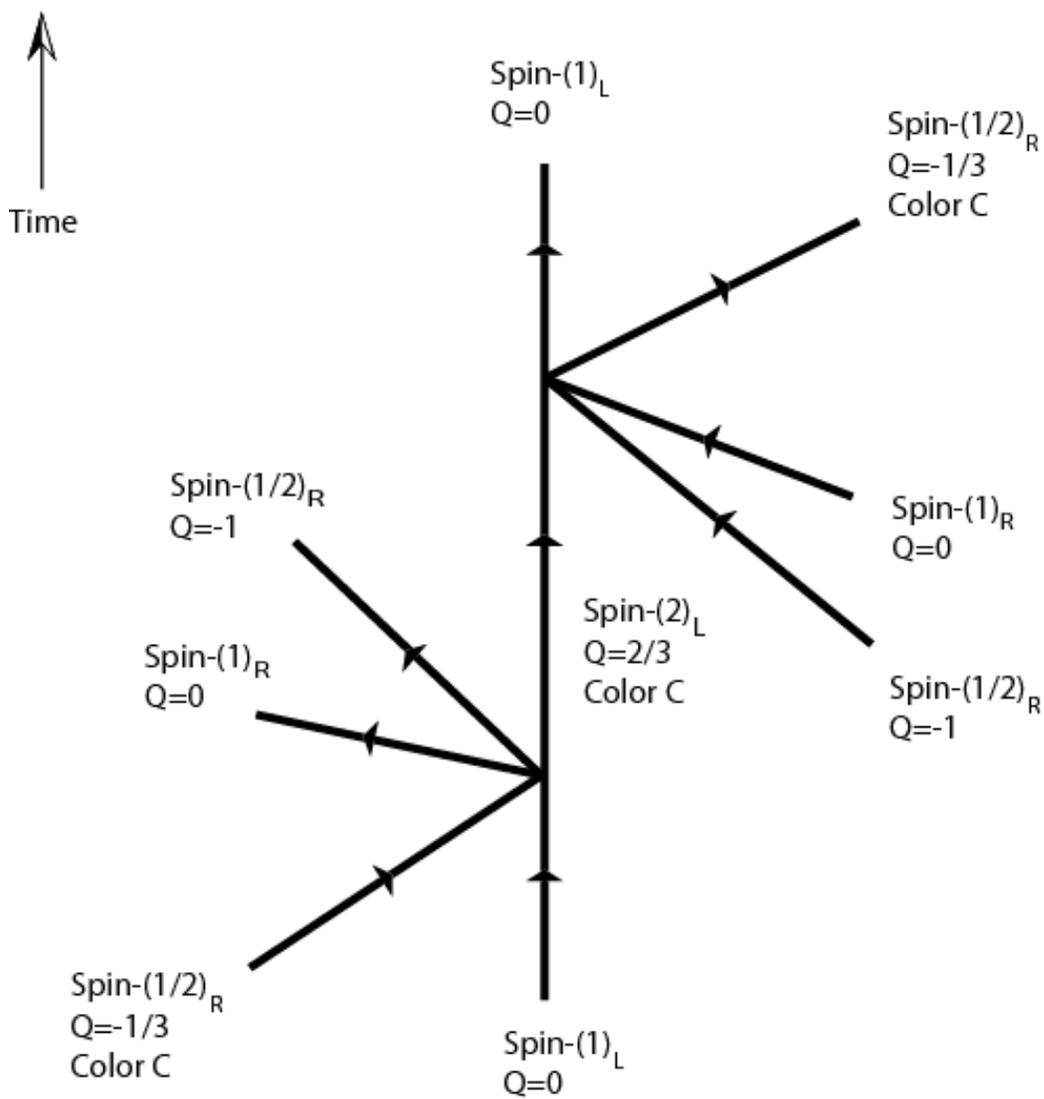

Figure 4